\documentclass[12pt]{article}
\usepackage{charter} % change font
\usepackage{fullpage}
\usepackage[hyphens]{url}     % allow hyphens in the url
\usepackage[hidelinks]{hyperref}       % hyperlinks
\usepackage{booktabs}       % professional-quality tables
\usepackage{amsfonts}       % blackboard math symbols
\usepackage{amssymb}        % varnothing
\usepackage{nicefrac}       % compact symbols for 1/2, etc.
\usepackage{microtype}      % microtypography
\usepackage{bm}				% italic bold
\usepackage{graphicx}		% figures
\usepackage{mathtools}		% for "coloneqq"
\usepackage{algorithm}
\usepackage{algpseudocode}
\usepackage{enumitem}		% to have letter enumerations
\usepackage{multirow}       % to generate tables
\usepackage{tabularx}	    % to generate tables
\usepackage{hhline}
\usepackage{arydshln}		% dotted lines in a table
\usepackage{siunitx}
\usepackage{caption}
\usepackage{subcaption}
\usepackage{wrapfig, capt-of}

\usepackage[table]{xcolor}
\usepackage{color}

\usepackage{makecell, multirow}

\definecolor{lightgreen}{rgb}{.9,1,.9}

\newcommand{\first}[1]{{\color{red} \textbf{#1}}}
\newcommand{\second}[1]{{\color{lightblue} \underline{#1}}}

\definecolor{lightgreen}{rgb}{.9,1,.9}
\definecolor{lightgray}{gray}{0.9}
\definecolor{lightred}{RGB}{255, 70, 70}
\definecolor{lightblue}{RGB}{70, 70, 255}
\definecolor{lightyellow}{RGB}{255, 255, 70}
\newcolumntype{L}[1]{>{\raggedright\arraybackslash}m{#1}}
\newcolumntype{C}[1]{>{\centering\arraybackslash}m{#1}}
\newcolumntype{R}[1]{>{\raggedleft\arraybackslash}m{#1}}

\def\defn{\,\coloneqq\,}
\def\lim{\mathop{\mathsf{lim}}} % limit

\def\xbf{{\mathbf{x}}}

\def\cbf{{\mathbf{c}}}
\def\xibf{\boldsymbol\xi}

\def\R{\mathbb{R}}

\def\Mcal{{\mathcal{M}}}

\def\Lcal{{\mathcal{L}}}

\def\Rcal{{\mathcal{R}}}
\def\Ucal{{\mathcal{U}}}
\def\Scal{{\mathcal{S}}}

\def\proposed{CURE}

\begin{document}

\title{Learning Cross-Video Neural Representations for High-Quality Frame Interpolation}

\author{
Wentao Shangguan$^{\ast, 2}$, Yu Sun$^{\ast, 1}$, \\
Weijie Gan$^{1}$,~and~Ulugbek~S.~Kamilov$^{1, 2}$\\
\emph{\footnotesize $^{1}$Department of Computer Science and Engineering,~Washington University in St.~Louis, MO 63130, USA}\\
\emph{\footnotesize $^{2}$Department of Electrical and Systems Engineering,~Washington University in St.~Louis, MO 63130, USA}\\
\footnotesize$^{\ast}$\emph{These authors have contributed equally to the work and are ordered alphabetically}
}

\date{}

\maketitle %% required

\begin{abstract}
This paper considers the problem of temporal video interpolation, where the goal is to synthesize a new video frame given its two neighbors. We propose \emph{\textbf{C}ross-Video Ne\textbf{u}ral \textbf{R}epres\textbf{e}ntation (\textbf{CURE})} as the first video interpolation method based on \emph{neural fields (NF)}. NF refers to the recent class of methods for neural representation of complex 3D scenes that has seen widespread success and application across computer vision. CURE represents the video as a continuous function parameterized by a coordinate-based neural network, whose inputs are the spatiotemporal coordinates and outputs are the corresponding RGB values.  \proposed~introduces a new architecture that conditions the neural network on the input frames for imposing space-time consistency in the synthesized video. This not only improves the final interpolation quality, but also enables \proposed~to learn a prior across multiple videos. Experimental evaluations show that \proposed~achieves the state-of-the-art performance on video interpolation on several benchmark datasets.
\end{abstract}

\section{Introduction}

The problem of temporal video interpolation seeks to synthesize new video frames given the observation of the neighboring frames. Video interpolation is crucial for many applications, including video compression~\cite{Lu.etal2017},  frame-rate conversion~\cite{Castagno.etal1996}, novel view synthesis~\cite{Flynn.etal2016}, and frame recovery~\cite{Wu.etal2015}. Traditional video interpolation approaches have considered variational optimization~\cite{Brox.etal2004}, nearest neighbor search~\cite{Chen2013}, or kernel filtering~\cite{Takeda.etal2008}. Recent work in the area has focused on deep learning (DL), where convolutional neural networks (CNNs) are trained to synthesize the desired frames. Two most widely-used classes of DL video interpolation methods are kernel-based~\cite{Niklaus.etal2017a,Niklaus.etal2017b,niklaus2021revisiting} or flow-based~\cite{Ren.etal2017,Ranjan.etal2017,Liu.etal2017video,Hui.etal2018,Xue.etal2019,Teed.etal2020,Bao.etal2021} methods. 

There has been considerable recent interest in the class of methods known as \emph{neural fields (NF)}~\cite{Sitzmann.etal2019,Park.etal2019}. NF seeks to represent varying physical quantities of spatial and temporal coordinates using fully-connected coordinate-based neural networks. The NF neural networks take spatial or temporal coordinates at the input and produce the corresponding physical quantity at the output. This coordinate-based representation frees NF methods from pre-defined pixel/voxel grids, thus enabling the learning of \emph{continuous} fields from discrete observations. For example, the popular \emph{neural radiance fields (NeRF)}~\cite{Mildenhall.etal2020} and its various extensions~\cite{Liu.etal2020voxel,Martin.etal2020} have achieved the state-of-the-art results in synthesizing novel views of complex 3D scenes from a set of 2D images. The NF approach has also been extended to the video setting by learning the representation of temporally-varying scenes from a set of videos~\cite{Li.etal2020neural,Du.etal2021,Xian.etal2020,Peng.etal2021,Li.etal2021neural}. Despite the recent activity, to the best of our knowledge, there is no NF method for temporal video interpolation that jointly addresses the lack of the camera-pose information, the need for test-time optimization, and the use of priors on the unknown video frames.

\begin{figure*}[t!]
\centering
\includegraphics[width=0.9\textwidth]{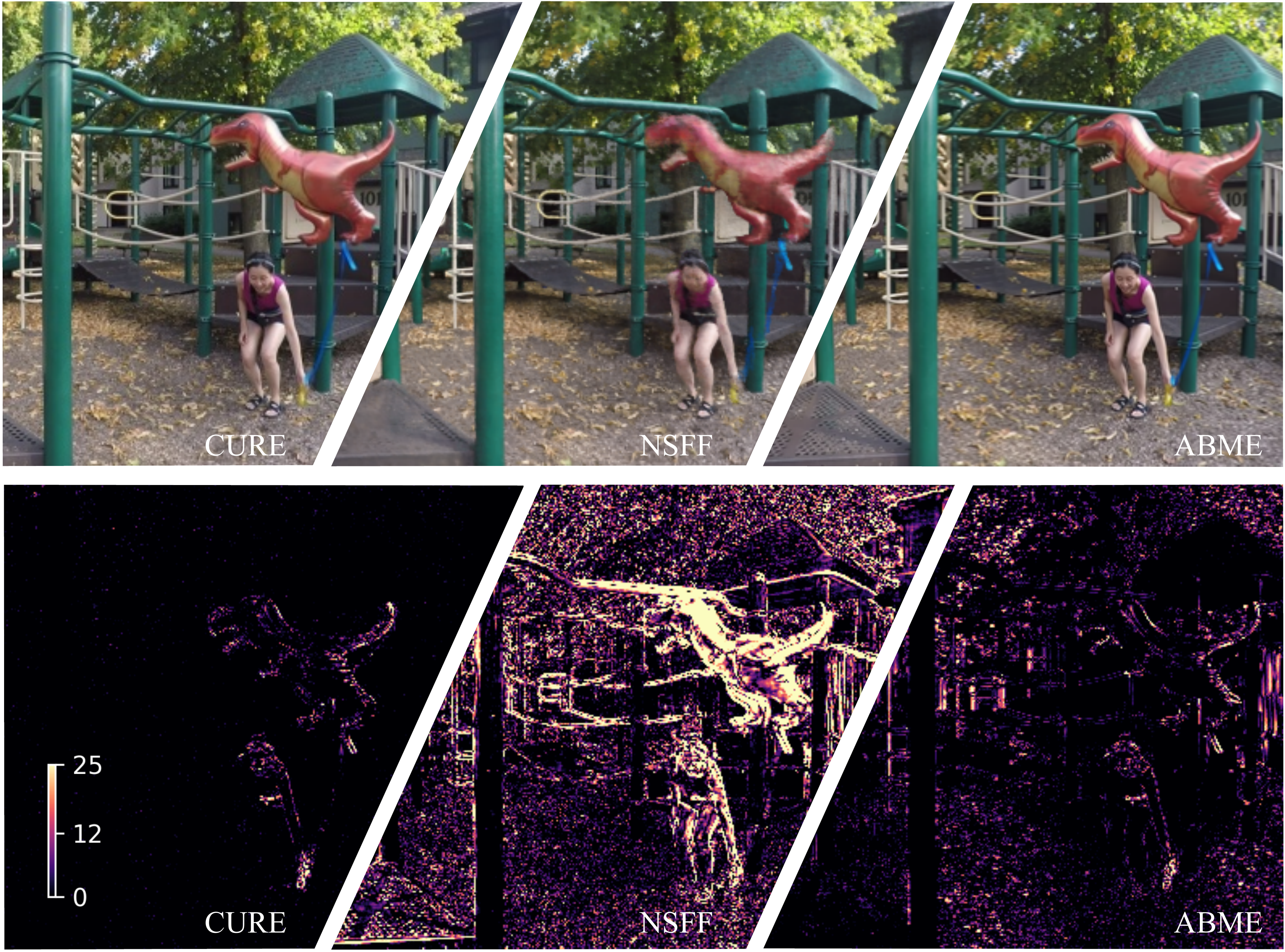}
\caption{Visual illustration of a frame interpolated by \proposed, NSFF~\cite{li2021neural}, and AMBE~\cite{park2021asymmetric}. NSFF and AMBE are the state-of-the-art methods for dynamic-scene view synthesis and video interpolation, respectively. Note how \proposed~substantially improves the visual quality and reduces the residual error in the second row.}
\label{fig:banner}
\end{figure*}

In this paper, we propose \emph{\textbf{C}ross-Video Ne\textbf{u}ral \textbf{R}epres\textbf{e}ntation (\textbf{CURE})} as a novel NF method for high-quality temporal video interpolation. Unlike traditional NF methods for representing dynamic scenes, \proposed~uses a deep neural network to directly represent a video as a continuous function of time and space. This conceptual change significantly simplifies the setup by circumventing the reliance on the camera pose information, which is usually difficult to estimate in certain applications such as handheld filming. The direct application of NF to video interpolation leads to poor performance due to the challenge of representing complex space-time variations from limited data. \proposed~addresses this issue by introducing a novel strategy based on conditioning the neural field on the input frames using our \emph{spatiotemporal encoding module (STEM)}.  For each queried pixel, STEM generates a local feature code that contains the spatial information around the pixel as well as the temporal information from the neighboring frames. STEM thus incorporates a space-time consistency prior into the representation network for mitigating visual artifacts. The architecture of conditioning also allows \proposed~to learn a video prior across different videos, enabling fast frame interpolation with a simple feed-forward pass. This is fundamentally different from the traditional NF approach of optimizing the weights of neural network for every new test video, thus making the inference computationally expensive.

\textbf{Contributions:} The technical contributions of this work are as follows:
\begin{enumerate}

\item 
We propose CURE as a novel NF approach for temporal video interpolation without any information on the camera pose. CURE is fundamentally different from the existing flow-based and kernel-based DL approaches.

\item
We develop a new architecture for \proposed~that leverages prior information over video frames, thus enforcing plausibility of the synthesized frames. The prior is obtained by conditioning the neural network on local feature codes encoding both temporal and spatial information extracted from the neighboring frames.

\item 
We perform extensive validation of \proposed~on several benchmark datasets, including UCF101~\cite{Soomro.etal2012}, Vimeo90K~\cite{Xue.etal2019}, SNU Films~\cite{choi2020channel}, Nvidia Dynamic Scene Dataset~\cite{yoon2020novel}, and Xiph4K. The empirical results show the state-of-the-art performance of \proposed~for temporal video interpolation as well as highlight its advantage over the current NF methods (see Fig.~\ref{fig:banner}). 

%We experimentally show that the video prior learned by \proposed~is general, making it applicable to video enhancement tasks without additional training on the corresponding datasets.
\end{enumerate}

\section{Related Work}
\label{sec:background}
In this section, we briefly review the DL-based methods for video interpolation and discuss several recent results in the area of neural fields.

\subsection{Deep Learning for Video Interpolation}
\label{sec:deeplearning}

A number of DL methods have been developed for video frame interpolation. 
%There have been considerable interests in leveraging deep learning for video frame interpolation.
Early work has proposed direct~\cite{Long.etal2016} and phase-base~\cite{Meyer.etal2018} methods that pioneered the usage of deep learning for the task. Recently, the research effort has shifted to flow-based and kernel-based methods due to their improved performance. In the following, we highlight some recent methods from these two classes.

\textbf{Flow-based methods}~\cite{Liu.etal2017video,Jiang.etal2018} rely on the optical flow to synthesize new temporal frames by warping the existing frames.
A common scheme consists of two modules, where the first module predicts the optical flow and the second module performs frame synthesis. A recent line of work has considered improvements due to the
estimation of task-oriented flow in a self-supervised manner~\cite{Xue.etal2019}, leveraging of multiple optical flow maps for characterizing bilateral motion~\cite{park2020bmbc,park2021asymmetric}, and adoption of recursive multi-scale flow learning~\cite{sim2021xvfi}.
Another line of work has investigated different warping strategies, such as warping both the original frames and their feature representations for extra information~\cite{Niklaus.etal2018} and softmax splatting for differentiable forward warping~\cite{niklaus2020softmax}.

\textbf{Kernel-based methods}~\cite{niklaus2021revisiting} view frame interpolation as convolution operations over local patches.
The idea was first introduced in AdaConv~\cite{Niklaus.etal2017a}, which uses spatial-adaptive kernels to capture both the local motion between the input frames and the coefficients for pixel synthesis in the interpolated frame.
The empirical success of AdaConv has spurred various follow-up work, including on the use of separable convolutional kernels~\cite{Niklaus.etal2017b}, jointly leveraging optical flow and adaptive convolutions~\cite{Bao.etal2021}, introducing spatiotemporal deformable convolutions~\cite{lee2020adacof,shi2021video}, and combining channel attention modules~\cite{choi2020channel}.
%However, a large kernel is necessary to handle strong motion, which inevitably requires a large amount of memory to process high-resolution images. 

The proposed \proposed~method represents a new approach based on continuous cross-video neural representation of videos using conditional NF. 

\subsection{Neural Fields}

NF has emerged as a popular vision paradigm with applications in 3D image and shape synthesis~\cite{Chen.etal2019,Park.etal2019,Sitzmann.etal2020}, tomography~\cite{Sun.etal2021a,Liu.Sun.etal2021b,shen2021nerp,lindell2021autoint}, audio processing~\cite{Sitzmann.etal2019}, and view rendering~\cite{Mildenhall.etal2020,Martin.etal2020,Zhang.etal2020}. 
A line of work related to this paper has considered applying NF for view synthesis of dynamic scenes, where the goal is to render a novel image of a temporally-varying scene from an arbitrary viewpoint. 
Examples include learning a moving human body from sparse multi-view videos~\cite{Peng.etal2021}, jointly learning deformable fields and scenes~\cite{li2021neural,Du.etal2021,Park.etal2020}, and 3D video synthesis from multi-view videos~\cite{Li.etal2021neural}. Although these method can be applied for video interpolation (e.g.\ by fixing the viewpoint in space), they are not designed for this task and usually lead to poor interpolation quality.
Another related line of work has applied NF to low-level video processing, including video denoising~\cite{Du.etal2021,Chen.etal2021}, video super-resolution~\cite{Du.etal2021}, and video compression~\cite{Chen.etal2021}. To the best of our knowledge, \proposed~is the first NF model designed for temporal frame interpolation.
% condition NF on encodings
%~\cite{Peng.etal2021,Yu.etal2021,ChenLiu.etal2021}

\section{Cross-Video Neural Representation}

We now present the technical details of \proposed. Fig.~\ref{fig:scheme} illustrates the overall structure of the proposed method. 
We begin by introducing the idea of representing a video segment as a function and then explain the architecture of STEM that infuses the consistency prior into CURE.

\begin{figure*}[!t]
%\begin{wrapfigure}{O}{0.5\textwidth}
\centering
\includegraphics[width=0.7\textwidth]{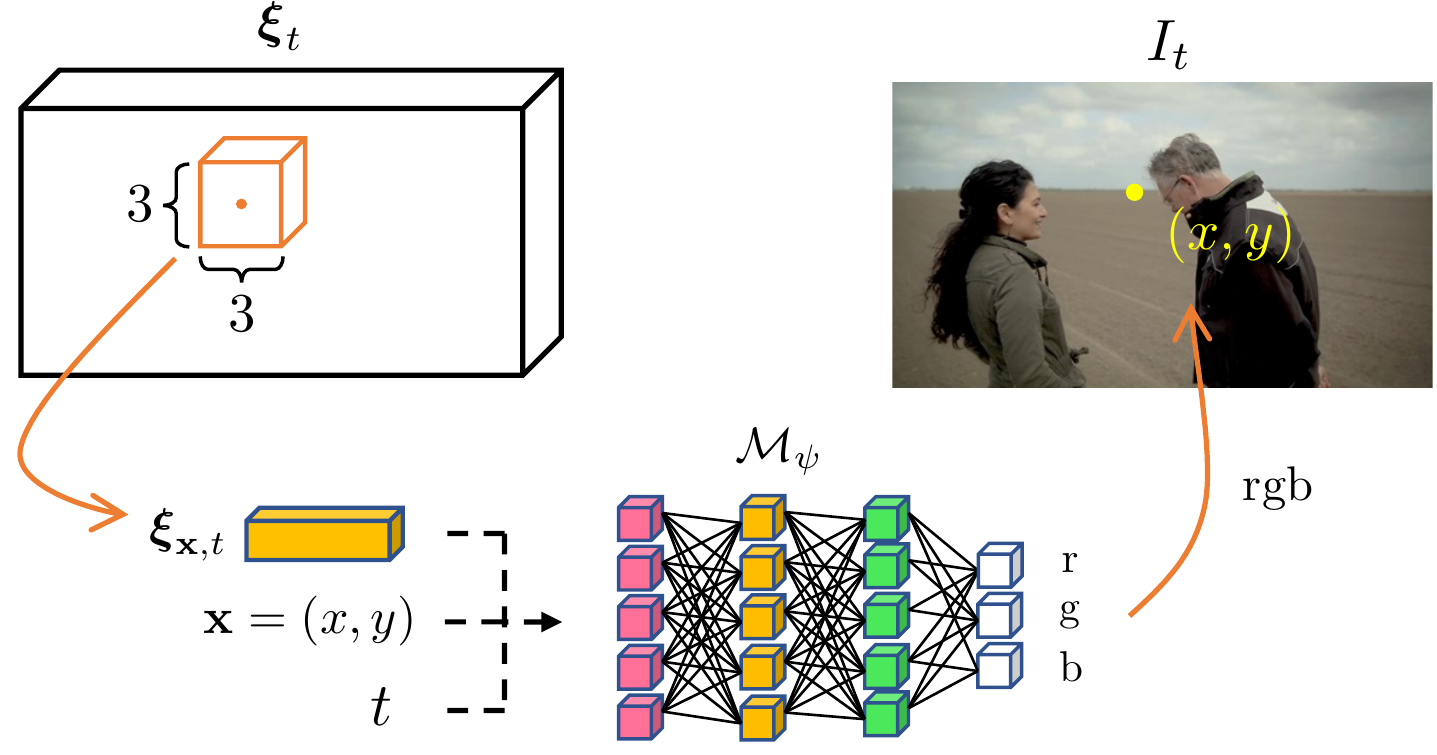}
\caption{\proposed~represents a video segment as a continuous vector-valued function, which maps $(\xbf, t, \xibf_{\xbf,t})$ to a single RGB color. The vector $\xibf_{\xbf,t}$ is the local feature code associated with the $3\times3$ region centered at $\xbf$ for the desired frame $t$. 
%We handle the boundary condition by using reflection.
}
\label{fig:concept}
\end{figure*}
%\end{wrapfigure}

\subsection{Representing Video As a Conditional Neural Field}

\proposed~represents a video segment between two observed frames as a continuous vector-valued function (see the visual illustration in Fig.~\ref{fig:concept}). The input to the function is the spatial pixel location $\xbf=(x,y)\in\R^2$, the relative time of the desired frame $0 \leq t \leq 1$, and an associated local feature code $\xibf_{\xbf,t}=\{\xi_{(x-1,y-1,t)}, \dots, \xi_{(x+1,y+1,t)}\}$ of the $3\times3$ region centered at $\xbf$. The output of the function is the single RGB value $\cbf=(r,g,b)\in\R^3$ at $\xbf$ in the predicted frame at time $t$. 
Here, $\xi_{(u,w, \tau)}$ denotes the latent code associated with location $(u,w)$ and time $\tau$, and we selected a $3\times3$ sub-region in order to impose pixel smoothness.
The formulation in \proposed~does not need any information on the camera pose, simplifying its use.  Additionally, the continuous formulation enables \proposed~to synthesize any frame given by the relative time $t$ (see Fig.~\ref{fig:time}).

\proposed~uses a neural network $\Mcal_\psi:(\xbf, t, \xibf_{\xbf, t}) \rightarrow \textbf{c}$ to approximate the coordinate-to-function mapping, where the $\psi$ denotes the network weights. The network contains 7 fully-connected layers with 256 neurons. Two skip connections are included in the second and fourth layers to concatenate the input vectors with the intermediate results. Fig.~\ref{fig:network}(c) shows the architectural details of CURE.
%\textcolor{red}{We observed that positional encoding is not beneficial in our case. Hence, we directly feed the raw coordinate $(\xbf,t)$ into the MLP.}
It is worth mentioning that as discussed in Sec.~\ref{sec:ablation} the exclusion of the spatiotemporal coordinates from CURE leads to suboptimal empirical performance.
\proposed~is trained over multiple videos for learning a general representation of videos. This is achieved by using the neighboring frames to generate the local feature codes. In the experiments, we trained \proposed~over a standard video interpolation dataset containing frame triplets. 

\subsection{Spatio-Temporal Encoding}
\label{sec:STEM}

Directly applying NF to video interpolation leads to poor results due to the lack of spatial and temporal consistency.
Fig.~\ref{fig:nvidia} illustrates the results obtained by the basic NeRF, which simply maps $(x,y,t)$ to the corresponding RGB value. The \emph{motivation} for spatio-temporal encoding is to provide additional prior information in both space and time to improve the neural representation performance.

\begin{figure*}[!t]
\centering
\includegraphics[width=0.9\textwidth]{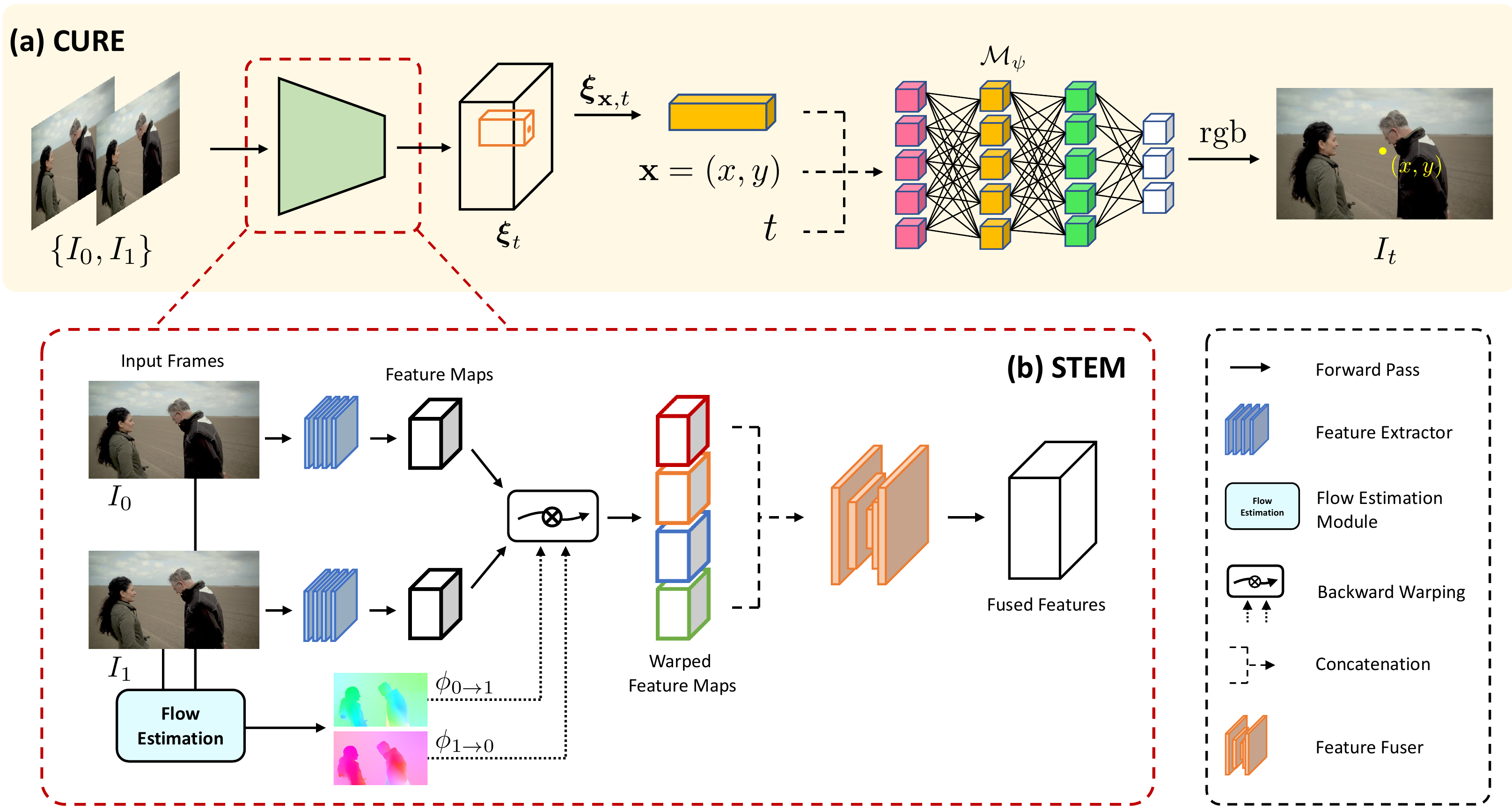}
\caption{(a) The overall architecture of \proposed. (b) The spatiotemporal encoding module (STEM) that generates $\xibf_t$ for the predicted frame $I_t$.}
\label{fig:scheme}
\end{figure*}

We design STEM to generate the local feature codes $\xibf_t=\{\xi_{(1,1,t)},...,\xi_{(H,W,t)}\}$ from the two neighboring frames $\{I_0,I_1\}$ of the desired frame $I_t$. Here, $H$ and $W$ denote pixel width and height of the video.
Fig.~\ref{fig:scheme}(b) illustrates the detailed architecture of STEM, which we denote as $\Scal_{\theta}: \{I_0,I_1\} \rightarrow \xibf_t$ for convenience.

\begin{figure*}[!t]
\centering
\includegraphics[width=0.88\textwidth]{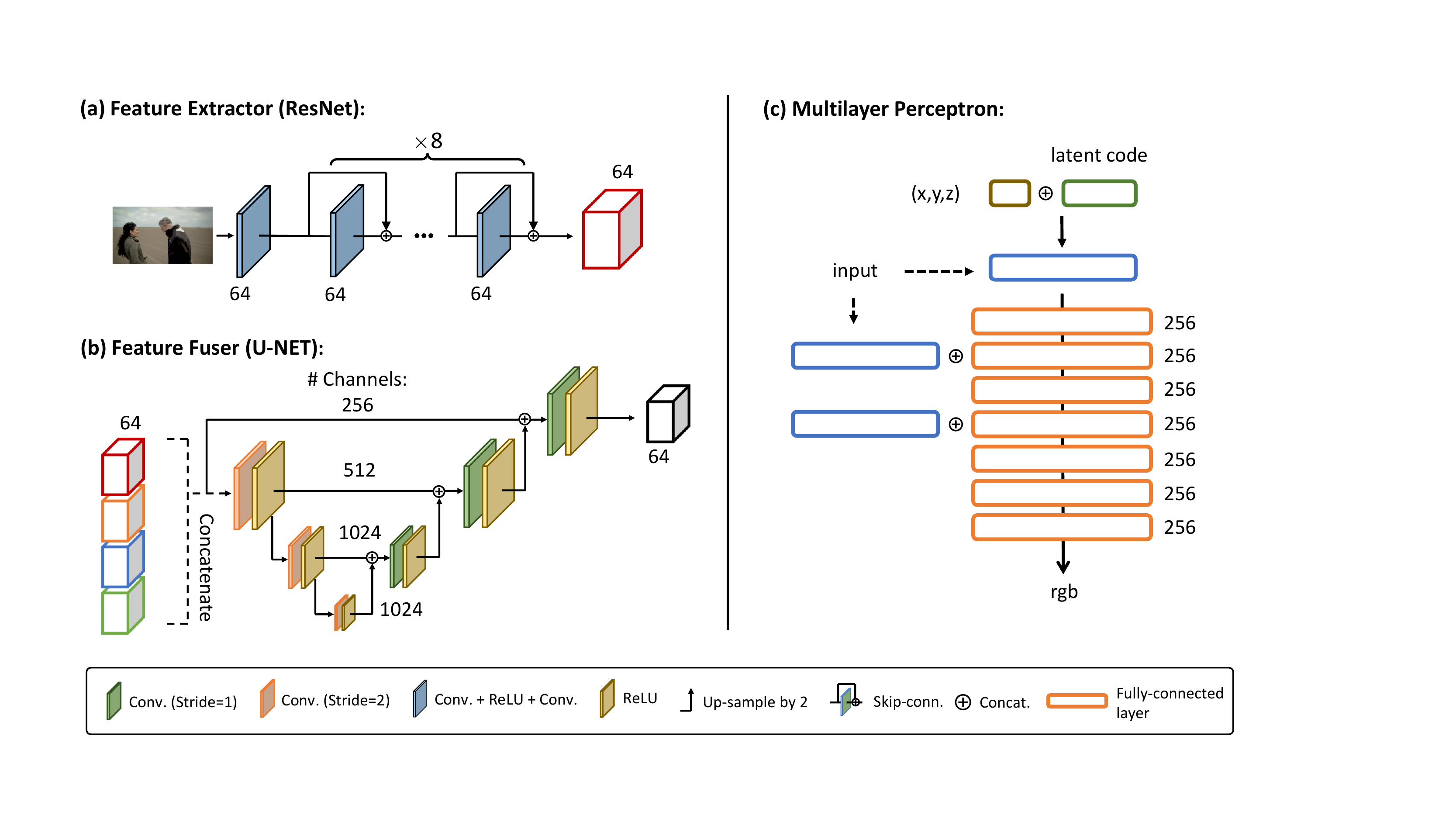}
\caption{Visual illustration of the network architectures used in \proposed.}
\label{fig:network}
\end{figure*}

\textbf{Feature Extraction.} STEM first extracts two individual $H \times W \times 64$ feature maps $\{F_0, F_1\}$ from the input frames by using a customized residual network (ResNet) $\Rcal: I\rightarrow F$,
\begin{equation}
F_0 = \Rcal(I_0),\quad F_1 = \Rcal(I_1).
\end{equation} 
The architecture of $\Rcal$ is shown in Fig.~\ref{fig:network}(a). 
Note that these feature maps are later used to form the latent feature map for frame $I_t$ via a two-step procedure: feature warping and fusion.

\textbf{Feature Warping.} Feature warping aims to interpolate the latent feature maps associated with time $t$ by warping $\{F_0, F_1\}$. 
This step ensures the temporal consistency between existing and predicted frames in the latent space.
To improve the interpolation accuracy, STEM adopts the bilateral motion approximation technique~\cite{park2020bmbc} to generate \emph{four} bilateral motion maps (fw: forward \& bw: backward) from the bidirectional optical flows $\phi_{0\rightarrow1}$ and $\phi_{1\rightarrow0}$,
\begin{subequations}
\label{eq:motion}
\begin{align}
&\phi_{t\rightarrow 0}^\text{fw} = (-t)\times\phi_{0 \rightarrow 1}, \\
&\phi_{t\rightarrow 1}^\text{fw} = (1-t)\times\phi_{0 \rightarrow 1}, \\
&\phi_{t\rightarrow 0}^\text{bw} = t\times\phi_{1 \rightarrow 0}, \\
&\phi_{t\rightarrow 1}^\text{bw} = -(1-t)\times\phi_{1 \rightarrow 0}.
\end{align}
\end{subequations}
Here, $\phi_{0\rightarrow1}$ and $\phi_{1\rightarrow0}$ are estimated by an off-the-shelf differentiable optical flow estimator. In our case, RAFT~\cite{Teed.etal2020} is used for flow estimation. 
STEM used the spatial transform network~\cite{Jaderberg2015} to warp the feature maps. In particular, we warp a target feature map $F_\text{tar}$ into a reference feature $F_\text{ref}$ by using the motion map from the reference to target
\begin{equation}
\label{eq:warping}
F_\text{ref}(\xbf)=F_\text{tar}(\xbf + \phi_{\text{ref}\rightarrow\text{tar}}(\xbf)) 
\end{equation}
By applying~\eqref{eq:warping} and~\eqref{eq:motion} to $\{F_0, F_1\}$, we can obtain four warped feature maps via cross combination
%$\{F^\text{fw}_{t,-1}, \\ F^\text{fw}_{t,+1}, F^\text{bw}_{t,-1}, F^\text{bw}_{t,+1}\}$ are given as
\begin{subequations}
\begin{align}
F_{0\rightarrow t}^\text{bw}(\xbf) = F_0(\xbf + \phi^\text{bw}_{t \rightarrow 0}(\xbf)), \\
F_{1\rightarrow t}^\text{bw}(\xbf) = F_1(\xbf + \phi^\text{bw}_{t \rightarrow 1}(\xbf)), \\
F_{0\rightarrow t}^\text{fw}(\xbf) = F_0(\xbf + \phi^\text{fw}_{t \rightarrow 0}(\xbf)), \\
F_{1\rightarrow t}^\text{fw}(\xbf) = F_1(\xbf + \phi^\text{fw}_{t \rightarrow 1}(\xbf)),
\end{align}
\end{subequations}
where a single warped feature map has the spatial dimension of $H \times W \times 64$. In feature warping, STEM adopted bilinear interpolation to handle off-the-grid pixels. 

\textbf{Feature Fusion.} To output the final local feature codes for $I_t$, a feature fusion step is then employed to fuse the warped feature maps into a single $H \times W \times 64$ feature map,
\begin{align}
\label{eq:fuse}
&\xibf_t = \Ucal(F_\text{concat})\\
\text{where}\quad&F_\text{concat} \defn F^\text{fw}_{0 \rightarrow t} \oplus F^\text{fw}_{1 \rightarrow t} \oplus F^\text{bw}_{0 \rightarrow t} \oplus F^\text{bw}_{1 \rightarrow t}.\nonumber
\end{align}
Here, $\oplus$ denotes feature concatenation along the channel dimension, and $\Ucal$ represents the fusion network. In STEM, we implemented a customized U-Net as $\Ucal$, which is shown in Fig~\ref{fig:network}(b).
%The bottleneck in \eqref{eq:fuse} can efficiently reduce the size of the latent code. 

%mix and distill the key temporal prior information for predicting $I_t$. 
%STEM used a customized U-Net, $\Ucal$, to perform feature fusion.
%The input of $\Ucal$ is a single feature map by concatenating the warped feature maps along the channel dimension
%The network combines four warped feature maps into one feature map
%The input of the network is four warped feature maps, and the o
%Fig.~\ref{fig:network} shows the detailed architecture. 

\subsection{Training of \proposed}

We jointly optimize the parameters of STEM and the coordinate-based neural network by minimizing the mean squared error (MSE) for each pixel location
\begin{align}
&\Lcal(\psi,\theta) = \|\Mcal_\psi(\xbf, t, \xibf_{\xbf,t}) - I_t(\xbf)\|_2^2 \quad\text{where}\quad\xibf_t = \Scal_\theta(I_0, I_1). 
\end{align}
The vector $\theta$ includes the weights of the feature extraction network $\Rcal$, feature fusion network $\Ucal$, and the optical flow estimator RAFT. We initialize RAFT with the default pre-trained weights, and set the RAFT iteration number to $50$.

We trained \proposed\ on the training dataset Vimeo90K~\cite{Xue.etal2019}. We performed data augmentation by vertically and horizontally flipping the video frames. \proposed~is trained by taking the start and end frames as input and predicting the middle frame at $t=0.5$.  Once trained, \proposed~can continuously interpolate frames at any $t\in(0,1)$ (see Fig.~\ref{fig:time}). We used Adam~\cite{Kingma.Ba2015} optimizer to optimize all network parameters for 70 epochs. We implemented a diminishing learning rate which decreases by 0.95 at every epoch. The training was performed on a machine equipped with an AMD Ryzen Threadripper 3960X 24-Core Processor and 2 NVIDIA GeForce RTX 3090 GPUs. It took roughly 3 weeks to train the model. 

\section{Experiments}
\label{sec:experiments}

We numerically validated \proposed~on several standard video interpolation datasets. Our experiments include comparisons with the state-of-the-art (SOTA) methods and ablation studies highlighting the key components of the proposed architecture.  

\subsection{Datasets}
We consider the following four datasets for numerical evaluation:
%\begin{itemize}

\textbf{Vimeo90K:} The Vimeo90K training dataset contains 51,312 frame triplets with $448\times256$ pixel resolution, and the validation dataset contains 3,782 triplets that we use for testing. We additionally increased the training data by extracting 91,701 triplets (i.e. $\{I_0, I_2, I_4\}$) from the septulets dataset.

\textbf{UCF101:} The UCF101~\cite{Soomro.etal2012} dataset contains realistic action videos collected from YouTube.
We used a subset collected by~\cite{Liu.etal2017video} for evaluation. The subset contains 379 triplets with $256\times256$ resolution.

\textbf{SNU-FILM:} The dataset~\cite{choi2020channel} contains $1,240$ triplets extracted from different videos with up to $1280\times720$ resolution. The triplets are categorized to \emph{easy}, \emph{medium}, \emph{hard}, and \emph{extreme}, based on the motion strength between frames.

\textbf{Xiph4K:} The dataset contains eight randomly-selected 4K videos downloaded from the Xiph website\footnote{https://media.xiph.org/video/derf/.}. For each video, we extracted the first fifty odd-numbered frames to form the testing triplets, all resized to $2048\times1080$ pixels.

\textbf{Nvidia Dynamic Scene (ND Scene):} ND Scene~\cite{yoon2020novel} is a video dataset that is commonly used for novel view synthesis. The dataset contains videos of eight scenes, recorded by a static camera rig with twelve GoPro cameras. We separate each video into triplets with a temporal sliding window of three frames. 

\textbf{X4K:} The X4K dataset~\cite{sim2021xvfi} contains eight 4K videos with 33 frames. 
We used this dataset for testing continuous multi-frame interpolation, with all videos resized to $2048\times1080$ pixels.
In the test, we used the first and last frames as input frames to predict seven middle frames, that is, $\{I_4, I_8, I_{12}, I_{15}, I_{20}, I_{24}, I_{28}\}$. 
%4,8,12,15,20,24,28

\textbf{Quantitative Metrics:} We use \emph{peak-signal-to-noise ratio (PSNR)} and \emph{structural similarity index (SSIM)} to quantify the performance of all methods. 

%%% Table 1 %%%

\begin{table*}[t!]
  \centering
  {
  \scriptsize
  \caption{Average PSNR/SSIM obtained by \proposed~and the current state-of-the-art (SOTA) methods on UCF101, Vimeo90K, and SNU-FILM, NDScene, and Xiph4K datasets. We use {\color{lightred}\textbf{bold}} and {\color{lightblue}\underline{underline}} to highlight the highest and second highest values, respectively. Note the superior results of CURE over SOTA methods.}
  \renewcommand\arraystretch{1.25}
  \begin{tabularx}{\textwidth}{C{.1\textwidth}C{.07\textwidth}C{.07\textwidth}C{.07\textwidth}C{.07\textwidth}C{.07\textwidth}C{.07\textwidth}C{.07\textwidth}C{.07\textwidth} | C{.07\textwidth}}
    \toprule
    \multirow{2}{*}{} & \multirowcell{2}{UCF\\101} & \multirowcell{2}{Vimeo\\90K} & \multicolumn{4}{c}{SNU-FILM} & \multirowcell{2}{ND\\Scene} & \multirow{2}{*}{Xiph4K} & \multirow{2}{*}{\textbf{Average}}      \\ 
    \cmidrule(lr){4-7} & & & Easy & Medium & Hard & Extreme & & & \\
    \midrule
	\rowcolor{lightgray}{SepConv~\cite{Niklaus.etal2017b} (ICCV17)}                     &  {27.97   0.9401}                     &  {27.32   0.9481}                     &  {30.33   0.9661}                     &  {27.33   0.9451}                     &  {22.80   0.8794}                     &  {18.06   0.7747} &  {24.38  0.9294}  &  {22.83  0.9748}  &  {25.13  0.9072}                     \\

     {ToFlow~\cite{Xue.etal2019} (IJCV19)}                      &  {34.58   0.9667}                     &  {33.73   0.9682}                      &  {39.08   0.9890}                     &  {34.39   0.9740}                     &  {28.44  0.9180}                     &  {23.39  0.8310}   &  {30.05  0.9466} &  {27.53  0.9850} &  {31.40  0.9361}                  \\

    %  {DAIN~\cite{}}                        & 34.99/0.968{\color{red}?}                     & 34.71/0.956{\color{red}?}                     & 39.73/0.9902                     & 35.46/0.9780                     & 30.17/0.9335                     & 25.09/0.8584                      \\

    \rowcolor{lightgray}{BMBC~\cite{park2020bmbc} (ECCV20)}                        &  {35.16   0.9689}                     &  {35.01   0.9764}                     & \first{39.90}   \second{0.9902}                    &  {35.31  0.9774}                     &  {29.33  0.9270}                     &  {23.92  0.8432}  &  \second{36.01} \second{0.9827} &  {28.03   0.9131} &  {32.83  0.9474}                   \\

     {CAIN~\cite{choi2020channel} (AAAI20)}                        &  {34.91   0.9690}                     &  {34.65  0.9730}                     &  {39.89  0.9900}                     &  {35.61  0.9776}                     &  {29.90  0.9292}                     &  {24.78  0.8507}  &  {34.81  0.9789}  &  {29.94  0.9184}  &  {33.06  0.9484}                   \\

     \rowcolor{lightgray}{RRIN~\cite{li2020video} (ICASSP20)}                       &  {34.93  0.9496}                     &  {35.22  0.9643}                     &  {39.31  0.9897}                     &  {35.23  0.9776}                     &  {29.79  0.9301}                     &  {24.59  0.8511}  &  {33.89  0.9750} &  {29.30  0.9232} &  {32.78  0.9451}                     \\

     {AdaCoF~\cite{lee2020adacof} (CVPR20)} &  {34.90  0.9680} &  {34.47  0.9730} &  {39.80  0.9900} &  {35.05  0.9754} &  {29.46  0.9244} &  {24.31  0.8439} &  {35.75  0.9822} &  {28.57  0.9093} &  {32.79  0.9458} \\

     \rowcolor{lightgray}{XVFI~\cite{sim2021xvfi} (ICCV21)} &  {35.09  0.9685} &  {35.07  0.9760} &  {39.76  0.9991} &  {35.12  0.9769} &  {29.30  0.9245} &  {23.98  0.8417} &  {35.76  0.9819} &  {28.46  0.9121} &  {32.82  0.9476}\\ 
    
     {ABME~\cite{park2021asymmetric} (ICCV21)}                        &  \first{35.38}   \second{0.9698}                     &  \first{36.18}   \first{0.9805}                    &  {39.59   0.9901}                     &  \second{35.77}   \second{0.9789}                     &  \second{30.58}  \second{0.9364}                     &  \second{25.42}   \first{0.8639} &  {33.17  0.9736} &  \second{30.74}  \second{0.9366} &  \second{33.35} \second{0.9537}                    \\

    \midrule
    {\proposed}           &  \second{35.36} \first{0.9705}   &  \second{35.73}   \second{0.9789}      &  \first{39.90}  \first{0.9910} &  \first{35.94}   \first{0.9797} &  \first{30.66}  \first{0.9373} &  \first{25.44}   \second{0.8638} &  \first{36.24}  \first{0.9839} &  \first{30.94}  \first{0.9389} &  \first{33.78}  \first{0.9555} \\
    \bottomrule

    \end{tabularx}
    \label{tb:main}
  }
\end{table*}

\begin{figure*}[t!]
  \centering
  \includegraphics[width=\textwidth]{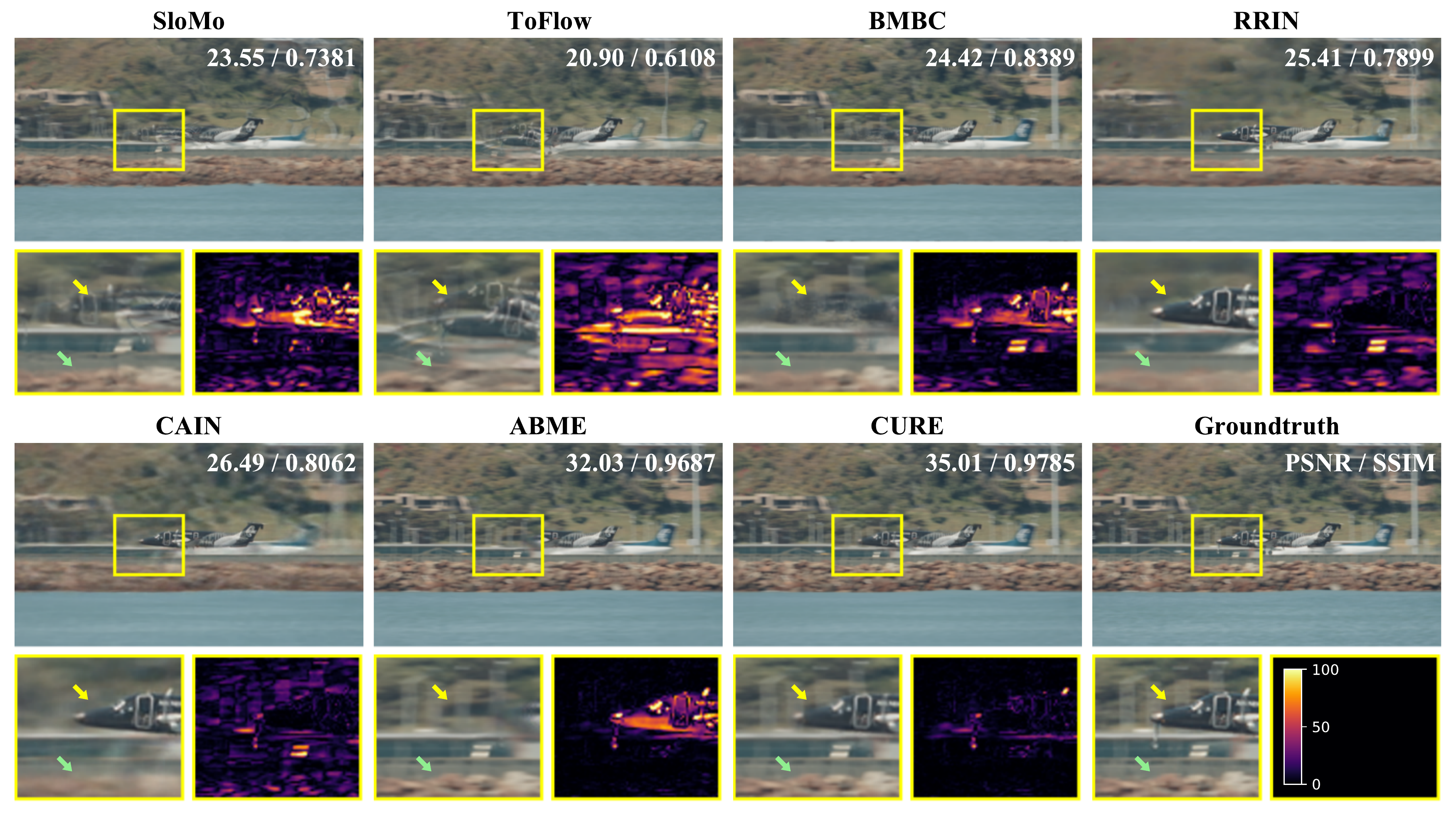}
  \caption{Visual comparison between \proposed~and current SOTA methods on an example from Vimeo90K. The visual differences are highlight by the yellow boxes paired with the residual error maps. Note how \proposed\ outperforms the SOTA methods in term of sharpness (see green arrows) and accuracy (yellow arrows).}
  \label{fig:vimeo90k}
\end{figure*}

\begin{figure*}[t!]
  \centering
  \begin{subfigure}[b]{\textwidth}
    \includegraphics[width=\textwidth]{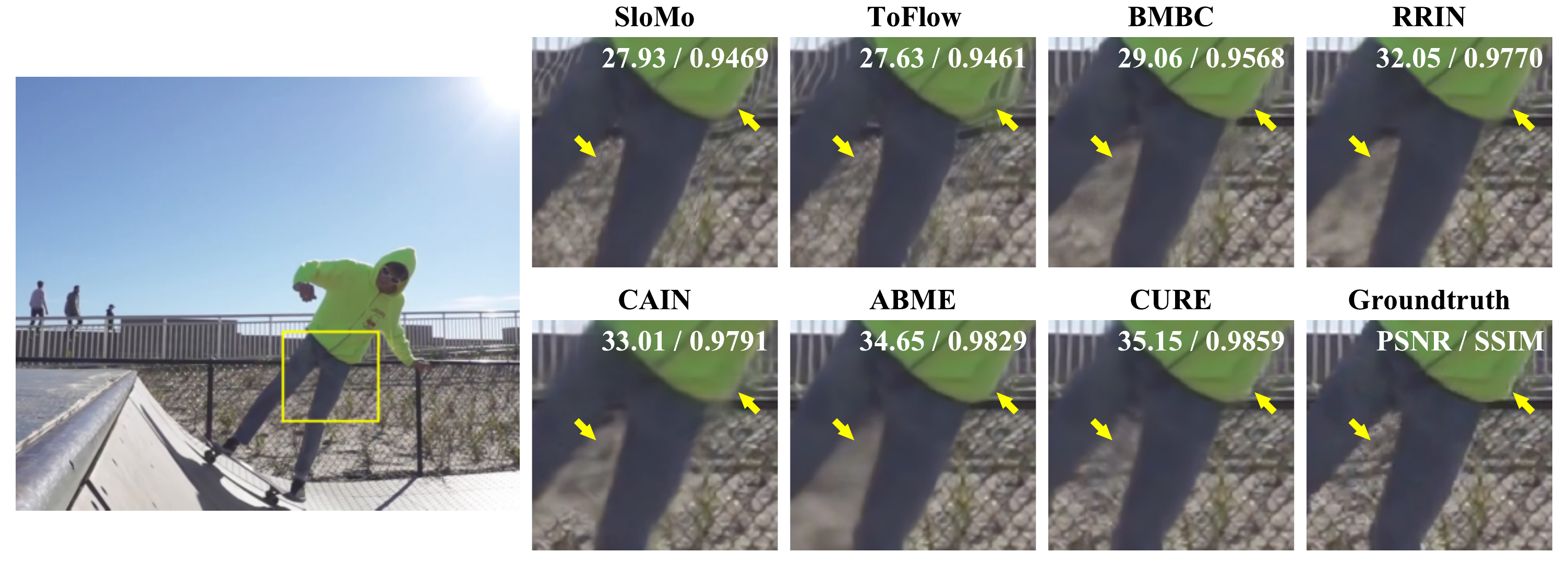}
    \end{subfigure}
  \begin{subfigure}[b]{\textwidth}
  \includegraphics[width=\textwidth]{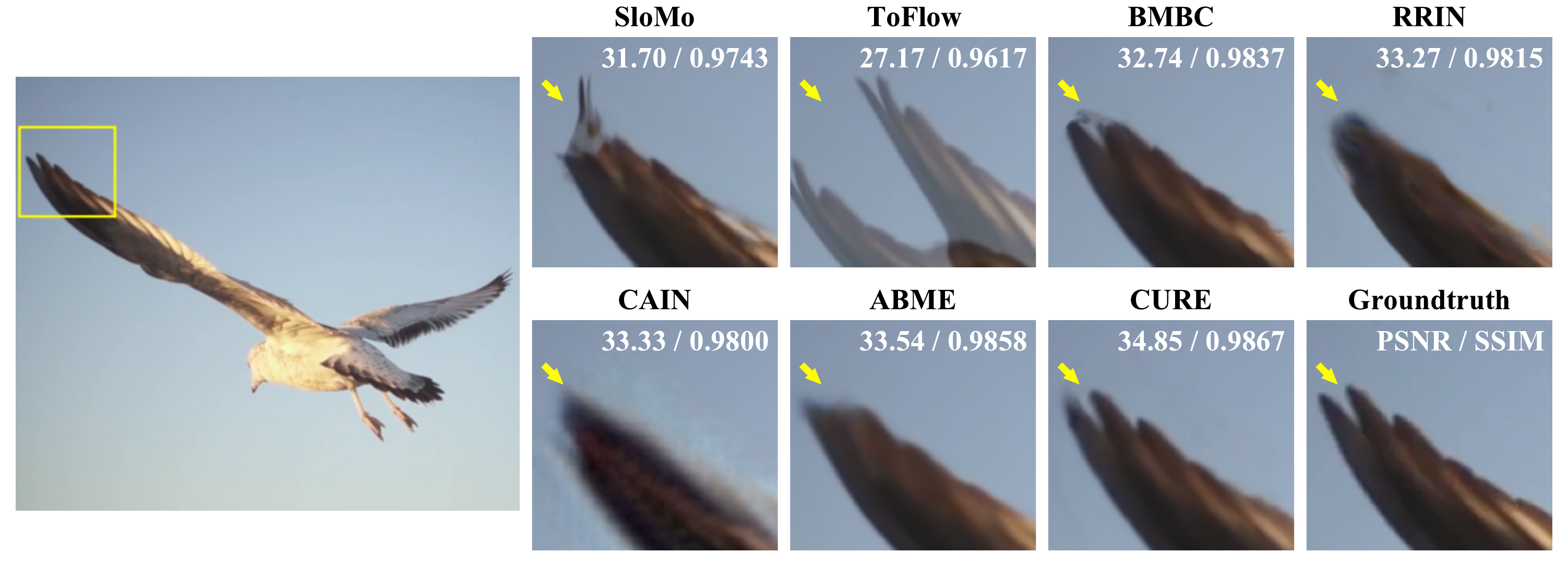}
  \end{subfigure}
  \caption{Visual comparison between \proposed~and current SOTA methods on two examples from SNU-FILM (hard). We highlight the differences by zooming in the regions with the presence of strong motion (the man's waist and the bird's wingtip). Note how \proposed~outperforms the SOTA methods by removing distortions and enhancing sharpness.}
  \label{fig:snu-film-hard}
\end{figure*}

\subsection{Performance Evaluation}

\textbf{Comparison with video interpolation methods.} Table~\ref{tb:main} summarizes the results of comparing \proposed~against several SOTA algorithms. The results are obtained by running the published code with the pre-trained weights. The average PSNR and SSIM values obtained by each algorithm on the considered datasets are presented.  Note how \proposed~outperforms all SOTA methods in terms of the PSNR/SSIM values averaged over all datasets.  For example, \proposed~outperforms ABME, which is the latest SOTA method, on most datasets and tied its performance (i.e.\ -0.1 dB in PSNR or -0.001 in SSIM) on the rest. 

%%% Table 1 %%%
\begin{wrapfigure}{O}{0.35\textwidth}
\centering
  \scriptsize
  \captionof{table}{Avg.\ PSNR/SSIM values obtained on X4K.}
  \renewcommand\arraystretch{1.5}
  \begin{tabularx}{0.31\textwidth}{C{.125\textwidth}|C{.16\textwidth}}
    \toprule
    \textbf{Methods} & \textbf{X4K}\\
    \rowcolor{lightgray}{SuperSloMo} & {19.58 / 0.7099}\\
    {ToFlow} & {22.33/0.7259} \\ 
    \rowcolor{lightgray}{BMBC} & {\second{24.28}  / \second{0.7777}} \\
    {XVFI} & {24.12/0.7566} \\
    \rowcolor{lightgray}{CURE} & {\first{30.05}  / \first{0.8998}}\\                       
    \bottomrule
    \end{tabularx}
    \label{tb:time}
\vspace{-5ex}
\end{wrapfigure} 

Fig.~\ref{fig:vimeo90k} visually compares \proposed~to several representative SOTA methods on an example from Vimeo90K. We highlight the visual differences by zooming in the yellow box and plotting the corresponding residual error map. In this comparison, SloMO, ToFlow, and BMBC reconstruct video frames with significant blurry and ghosting artifacts, failing to capture the shape of the airplane front (see yellow arrows). While RRIN and CAIN successfully recover the airplane front, they completely blur the mountain in the background (see green arrows). Note how \proposed~clearly reconstructs both the moving airplane as well as the background mountain with fine details. The qualitative visual improvements are also corroborated by the higher quantitative values and lower residual errors.

Fig.~\ref{fig:snu-film-hard} provides a visual results on two videos from SNU-FILM (hard). These two examples were selected due to the fast motion of the recorded objects, which makes the frame interpolation challenging. ToFlow completely fails to reconstruct the bird's wingtip in the second video. Other algorithms, such as BMBC, RRIN, CAIN, and AMBE, produce blurry frames that smooth out the details on the feathers. \proposed~successfully handles the motion by reconstructing a sharp frame. Note the visual similarity between the frame synthesized by \proposed~and the groundtruth. Similar results can be observed in the comparison on the first video of a man skateboarding, where \proposed~significantly outperforms other algorithms.

\begin{figure*}[t!]
  \centering
  \includegraphics[width=\textwidth]{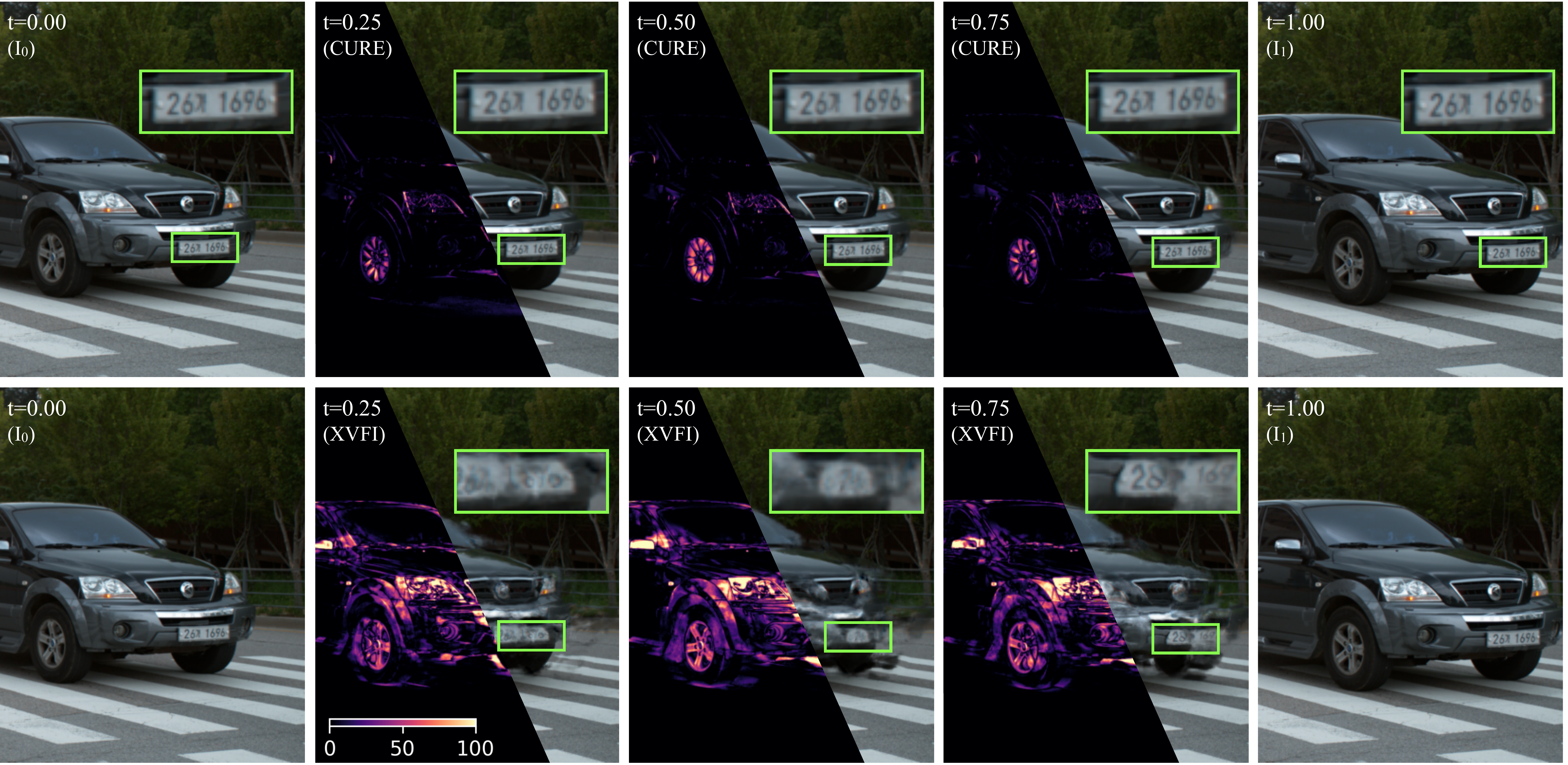}
  \caption{Visual demonstration of the continuous interpolation by \proposed~on an example from the X4K dataset. Only two frames $\{I_0,I_1\}$ are used for predicting several middle frames at temporal locations $t\in\{0.25,0.5,0.75\}$. Note the sharp textural details preserved by \proposed, but lost in the results by XVFI.}
  \label{fig:time}
\end{figure*}

Fig.~\ref{fig:time} highlights \proposed's ability to perform continuous interpolation on a challenging video from X4K. We include the result by XVFI as the baseline. Each figure plots both image and the corresponding residual error map with respect to the groundtruth. The visual differences on the textural details are highlighted in the green boxes. Note that both \proposed~and XVFI were only trained to predict the frame at $t=0.5$ on the Vimeo90K dataset.
%is only trained to predict the frame at $t=0.5$, while XVFI is trained to predict the frames at random temporal locations $t\in(0,1)$. 
\proposed~produces high-quality and accurate frames for every temporal location, significantly outperforming XVFI in terms of both visual quality and precision. Note how \proposed~recovers the vehicle license plate that is completely distorted by XVFI. Corresponding quantitative results are summarized in Table~\ref{tb:time}, corroborating the excellent performance of \proposed.

\textbf{Comparison with NF methods.}
We compared \proposed\ with two different NF algorithms: (1) vanilla NF (V-NF) that directly learns a coordinate-based neural network to map $(\xbf,t)$ to the corresponding color pixel, and (2) NSFF~\cite{li2021neural} that learns the dynamic scene in the video and applies view synthesis to generate new frames. Both algorithms need to re-optimize their weights for every test video/scene, while \proposed~learns a general video prior. We implemented V-NF by adopting the neural network architecture of NeRF and setting the positional encoding $L$ to 10 and 2 for $(x,y)$ and $t$, respectively. We trained V-NF by using the even-number frames. For NSFF, we downloaded the pre-trained models for testing. Note that these comparisons are provided for completeness only; the performance of these NF algorithms on video interpolation should not be interpreted as an indication of their performance on their original tasks. 

\begin{figure*}[t!]
  \centering
  \includegraphics[width=\textwidth]{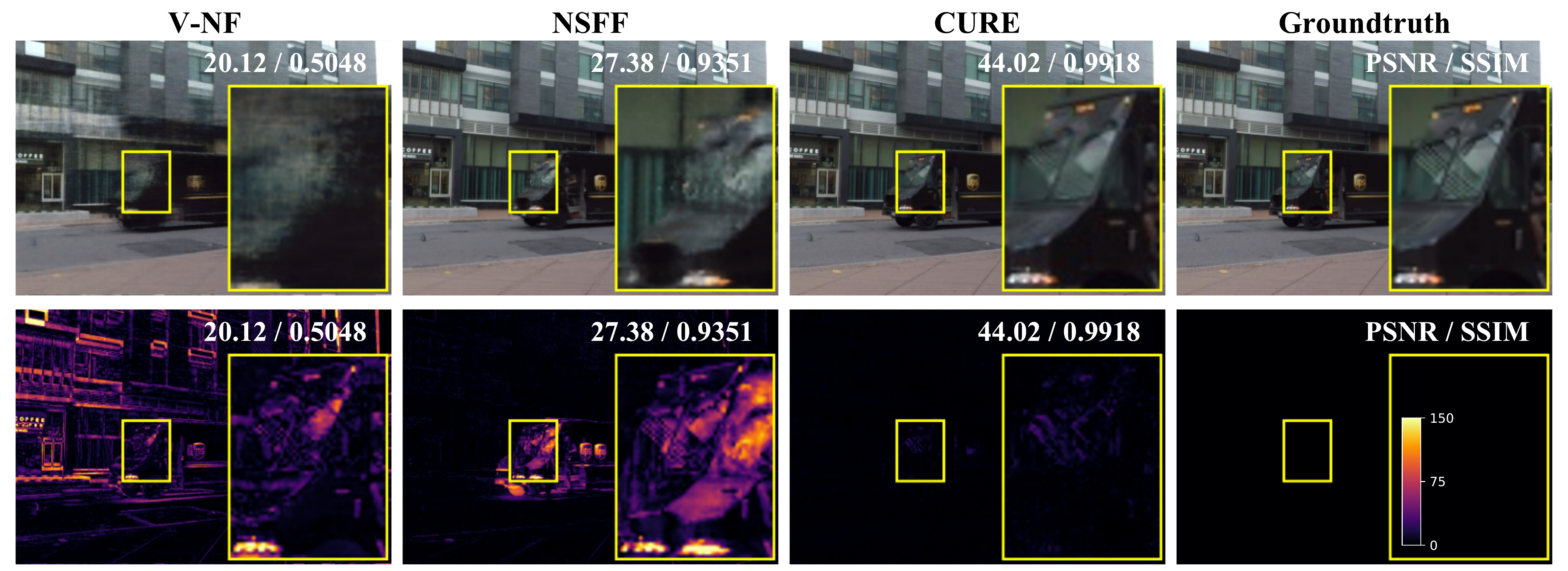}
  \caption{Visual comparison between \proposed, vanilla-NeRF, and NSFF methods on ND Scene. The first row shows the images, and the second row shows the corresponding residual error map. We highlighted the visual difference in the yellow box. \proposed~achieves substantially higher PSNR (about 17 dB) over NSFF.}
  \label{fig:nvidia}
\end{figure*}

We evaluated the comparison on ND Scene. For each scene, we randomly selected one camera video from the videos captured by twelve cameras.  We resized each video to accommodate the setup used by the pre-trained models of NSFF.
Table~\ref{tb:nvidia} summarizes the details and quantitative results on each test video.  As illustrated, \proposed~significantly outperforms V-NF and NSFF in every test scenario, and the PSNR improvement is usually over 5 dB. Fig.~\ref{fig:nvidia} presents a visual comparison of the \emph{truck} video to better highlight the difference.Without any consistency constraint, V-NF faithfully recovers the general content in the frame, but loses the feature details and suffers strong gridding artifacts. For example, the residual map highlights the missing edges in the interpolated frame in V-NF. On the other hand, NSFF generates a visually pleasing frame with faithful details. However, the problem of NSFF is that it predicts the wrong position of the moving truck, which significantly reduces the PSNR values. \proposed\ avoids all these problems and interpolates a high-quality frame with both fine details and correct position of the truck.  These results highlight the contribution of \proposed~to the research on NF-based video frame interpolation.

%%% Table %%%

\begin{table*}[t!]
  \centering
  {\scriptsize
  \caption{Quantitative (PSNR/SSIM) comparisons of different video representation algorithms on the ND Scene dataset. This table highlights that CURE provides better video representation by learning a general prior across videos.}
  \renewcommand\arraystretch{1.5}
  \setlength{\tabcolsep}{5pt}
  \begin{tabular}{cccccc}
    \toprule
    \multirow{2}{*}{Dataset}        & \multirow{2}{*}{Camera Post} & \multirow{2}{*}{Resolution} & \multicolumn{3}{c}{Algorithms} \\
    \cmidrule(lr){4-6}
    & & & V-NF       & NSFF~\cite{li2021neural}         & CURE                  \\
    \midrule
    \rowcolor{lightgray} Ballon1-2      & 1    & $540 \times 288$       & 10.75/0.4582 & 24.58/0.8980 & \textbf{31.05/0.9742} \\
    Dynamic Face-2 & 3    & $547 \times 288$       & 23.09/0.8317 & 27.13/0.9608 & \textbf{40.81/0.9930} \\
    \rowcolor{lightgray}Jumping        & 5    & $540 \times 288$       & 26.07/0.8099 & 27.19/0.9281 & \textbf{31.23/0.9624} \\
    Playground     & 2    & $573 \times 288$       & 22.45/0.7222 & 24.64/0.8912 & \textbf{36.15/0.9938} \\
    \rowcolor{lightgray}Skating-2      & 4    & $541 \times 288$       & 28.44/0.8648 & 34.05/0.9782 & \textbf{39.67/0.9899} \\
    Truck-2        & 9    & $542 \times 288$       & 28.37/0.8075 & 34.74/0.9751 & \textbf{39.78/0.9887} \\
    \rowcolor{lightgray}Umbrella       & 11   & $543 \times 288$       & 23.91/0.5876 & 23.91/0.8441 & \textbf{39.66/0.9880} \\
    \bottomrule
    \end{tabular}
    \label{tb:nvidia}
  }
\end{table*}

\subsection{Ablation Studies}
\label{sec:ablation}
We highlighted the key components of the proposed model by conducting several ablation experiments. 
%As the importance of STEM has already been shown in the comparison with NF methods,  
We focus on showing the effectiveness of the coordinate-based neural representation by considering two baseline variants: 

\textbf{\proposed-w/o-Rep}: This method directly trains the STEM network as a frame interpolator by adding an additional head to predict the RGB images. We design \proposed-w/o-Rep to show the importance of the NF module.

\textbf{\proposed-w/o-Crd}: This method simplifies \proposed~by removing the input of the spatiotemporal coordinate $(\xbf,t)$. We design \proposed-w/o-Crd to show the importance of the coordinate indexing for synthesizing the desired RGB values.

\begin{figure*}[t!]
  \centering
  \includegraphics[width=\textwidth]{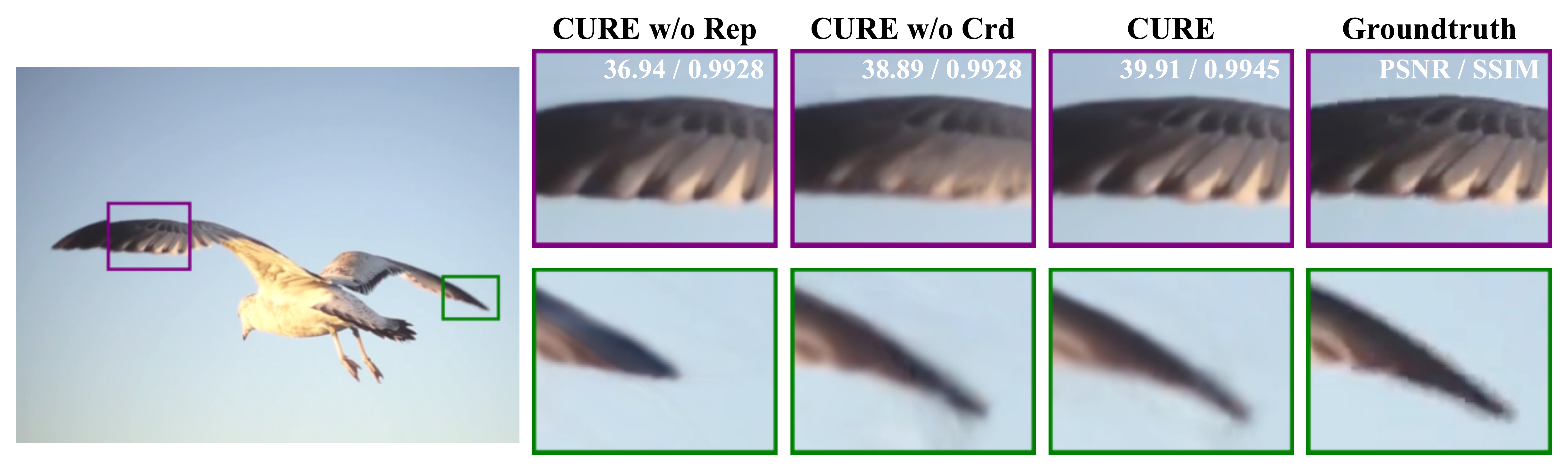}
  \caption{Visual illustration of the importance of the various components of \proposed. The green and yellow boxes highlight the improvements in terms of accuracy and sharpness due to the full \proposed~compared to its ablated variants.}
  \label{fig:abalation}
\end{figure*}

Fig.~\ref{fig:abalation} visually compares \proposed~with its two baseline variants on an example from SNU (hard). Without neural representation, \proposed-w/o-Rep predicts the wrong location of the bird's wingtip shown in the green box, losing enough representation power to properly represent the video. On the other hand, \proposed-w/o-Crd produces a blurry image that loses the texture details by missing the exact spatiotemporal coordinate. Table~\ref{tb:ablation} summarizes the numerical values on all the considered datasets. The results clearly show that \proposed~achieves the best results by jointly leveraging STEM and coordinate-based representation.
%The testing videos are from SNU-FILM (Medium) datasets. 
%Without the spatiotemporal coordinate to provide information on the pixel location, \emph{\proposed\ w/o C} suffers from blurry artifacts in its reconstructed video frame (see plumage details in purple bounding boxes). 
%While \emph{\proposed\ w/o MC} can preserve sharp details, its interpolated video frame is inconsistent with the ground-truth (see shape of plumage in green bounding box). 
%Overall, the original \proposed\ can achieve the best visual performance.

\begin{table*}[t!]
  \centering
  {
  \scriptsize
  \caption{Average PSNR/SSIM obtained by the full \proposed~and its ablated variants.}
  \renewcommand\arraystretch{1.35}
  \begin{tabularx}{\textwidth}{C{.1\textwidth}C{.07\textwidth}C{.07\textwidth}C{.07\textwidth}C{.07\textwidth}C{.07\textwidth}C{.07\textwidth}C{.07\textwidth}C{.07\textwidth} | C{.07\textwidth}}
    \toprule
    \multirow{2}{*}{} & \multirowcell{2}{UCF\\101} & \multirowcell{2}{Vimeo\\90K} & \multicolumn{4}{c}{SNU-FILM} & \multirowcell{2}{ND\\Scene} & \multirow{2}{*}{Xiph4K} & \multirow{2}{*}{\textbf{Average}}      \\ 
    \cmidrule(lr){4-7} & & & Easy & Medium & Hard & Extreme & & & \\
    \midrule
    \rowcolor{lightgray} {\proposed\ w/o Rep} & {34.80 0.9682} & {35.15 0.9769} & {39.82 0.9903} & {35.71 0.9787} & {30.36 0.9350} & {25.14 0.8601} & {35.13 0.9802} & {30.45 0.9359} & {33.32 0.9530} \\
    {\proposed\ w/o Crd} &  {35.11 0.9689} &  {35.25 0.9772} &  {39.76 0.9900} &  {35.63 0.9784} &  {30.39 0.9350} &  {25.26 0.8609} &  {35.68 0.9816} &  {30.52 0.9341} &  {33.45 0.9531} \\
    \midrule
    {\proposed}           &  \textbf{35.36} \textbf{0.9705}   &  \textbf{35.73}   \textbf{0.9789}      &  \textbf{39.90}  \textbf{0.9910} &  \textbf{35.94}   \textbf{0.9797} &  \textbf{30.66}  \textbf{0.9373} &  \textbf{25.44}  \textbf{0.8638} &  \textbf{36.24}   \textbf{0.9839} &  \textbf{30.94}  \textbf{0.9389} &  \textbf{33.78}  \textbf{0.9555} \\
    \bottomrule

	\end{tabularx}
  \label{tb:ablation}
  }
\end{table*}

\section{Conclusion}
This work proposes a novel temporal video frame interpolation algorithm, CURE, which is based on the recent class of techniques known as neural fields (NF). The key idea of \proposed~is to represent video segment between two observed frames as a continuous vector-valued function of the spatiotemporal coordinates, which is parameterized by a fully-connected coordinate-based neural network. Unlike the traditional NF variants, CURE enforces space-time consistency in the synthesized frames by conditioning its NF on the information from the neighboring frames. This conditioning enables CURE to learn a prior across different videos. Experimental evaluations clearly show that the CURE outperforms the state-of-the-art methods on five benchmark datasets. The comparisons with the traditional NF methods and with the ablated variants of CURE highlight the synergistic effect of various components of the full CURE. 

\section*{Acknowledgement}
This work was supported by the NSF award CCF-2043134.

\end{document}